\documentclass[a4paper]{jpconf}
\usepackage{graphicx}
\usepackage{float}
\usepackage{amsmath}

\def\Journal#1#2#3#4{{#4} {#1} {\bf #2}, #3}
\def\NIMA{{\em Nucl. Instrum. Methods} A}
\def\NPB{{\em Nucl. Phys.} B}
\def\PLB{{\em Phys. Lett.}  B}
\def\PRL{\em Phys. Rev. Lett.}
\def\PRD{{\em Phys. Rev.} D}
\def\JHEP{\em J. High Energy Phys.}

\begin{document}
\title{Gluon polarization measurements from longitudinally polarized proton-proton collisions at STAR}

\author{Zilong Chang, for the STAR Collaboration}

\address{Brookhaven National Laboratory, Upton, NY 11973}

\ead{zchang@bnl.gov}

\begin{abstract}
The gluon polarization contribution to the proton spin is an integral part to solve the longstanding proton spin puzzle. At the Relativistic Heavy Ion Collider (RHIC), the STAR experiment has measured jets produced in mid-pseudo-rapidity, $|\eta| < $ 1.0, and full azimuth, $\phi$, from longitudinally polarized $pp$ collisions to study the gluon polarization in the proton.  At center of mass energies $\sqrt{s} =$ 200 and 510 GeV, jet production is dominated by hard QCD scattering processes such as gluon-gluon ($gg$) and quark-gluon ($qg$), thus making the longitudinal double-spin asymmetry ($A_{LL}$) sensitive to the gluon polarization. Early STAR inclusive jet $A_{LL}$ results at $\sqrt{s} = $ 200 GeV provided the first evidence of the non-zero gluon polarization at momentum fraction $x > $ 0.05. The higher center of mass energy $\sqrt{s} = 510$ GeV allows to explore the gluon polarization as low as $x \sim$ 0.015. In this talk we will present the recent STAR inclusive jet and dijet $A_{LL}$ results at $\sqrt{s} = 510$ GeV, and discuss the relevant new analysis techniques for the estimation of trigger bias and reconstruction uncertainty, the underlying event correction on the jet energy and its effect on jet $A_{LL}$. Dijet results are shown for different topologies in regions of pseudo-rapidity, effectively scanning the $x$-dependence of the gluon polarization.
\end{abstract}

\section{Introduction}
Early deep inelastic scattering (DIS) experiments in the 1980s showed that quarks inside the proton make only a small contribution to its total spin \cite{emc1988}. Where the rest of the proton spin comes from has been an outstanding problem awaiting to be explored. Theorists introduced parton distribution functions, PDFs, to describe the probability of a parton with momentum fraction $x$ encountering a probe at energy scale, $Q^2$, $f(x, Q^2)$. Jaffe and Manohar proposed not only do quarks contribute to the proton spin, but also gluons and the orbital angular momentum of quarks and gluons \cite{jaff1990}. However the kinematics space in $x-Q^2$ covered by the polarized fixed target experiments through DIS processes only provide limited constraints on the gluon polarization inside the proton \cite{lss10}.

Different from DIS experiments, a polarized hadron-hadron collider at high center-of-mass energy, $\sqrt{s}$, such as the Relativistic Heavy Ion Collider (RHIC) \cite{rhic,rhicdesign,rhicpol} can provide
direct access to gluon polarization inside the proton. At RHIC, either transversely or longitudinally polarized proton beams collide at both $\sqrt{s} =$ 200 and 510 GeV. To explore the gluon polarizations, we measure the longitudinal double-spin asymmetry, $A_{LL}$, for jets, defined as the fractional difference of the jet cross sections when beams have the same and the opposite helicities. The $A_{LL}$ can be expressed as the sum of convolutions of the polarized PDFs and partonic longitudinal double-spin asymmetry $\hat{a}_{LL}$ over all possible partonic processes. The next-to-leading order (NLO) perturbative quantum chromodynamics (pQCD) calculations show that the $qg$ and $gg$ processes dominate jet production at both $\sqrt{s} =$ 200 and 510 GeV \cite{nlo2012}. Both $qg$ and $gg$ processes have sizable $\hat{a}_{LL}$ \cite{nloaLL2007}, therefore jet $A_{LL}$ are sensitive to gluon polarizations. The same applies to $A_{LL}$ measurements for hadrons, for example $\pi^0$. Given beam polarizations, $P_1$ and $P_2$, and relative luminosities $R = \frac{L^{++}}{L^{--}}$, $A_{LL}$ is defined experimentally as:

   \begin{align}
    A_{LL} = \frac{1}{P_1P_2}\frac{N_{++}-R N_{+-}}{N_{++}+RN_{+-}}.
  \end{align}

\section{Inclusive jet and dijet $A_{LL}$ measurements at STAR}

Solenoidal Tracker at RHIC (STAR) \cite{star} has published a series of inclusive jet and dijet $A_{LL}$ results at $\sqrt{s} = $ 200 GeV \cite{run9aLL, run9bdj, run9edj}. The inclusive jets with transverse momentum, $p_T$, and pseudo-rapidity, $\eta$, sample the scattering parton $x \approx x_T e^{\pm \eta}$, where $x_T = \frac{2p_T}{\sqrt{s}}$. The dijets are able to unfold the initial kinematics $x_1$, $x_2$, and the scattering angle in the parton scattering rest frame, $cos\theta^{\ast}$, as in Equations \ref{eq:x1}, \ref{eq:x2} and \ref{eq:theta}. Since the kinematics of the two scattering partons are simultaneously determined by dijet kinematics, it constrains the shape of polarized gluon distribution function, $\Delta g(x)$, as function of $x$. At $\sqrt{s} = $ 200 GeV, jets are sensitive to $\Delta g(x)$ at $x$ as low as 0.05 when $|\eta|  < $ 1.0. The new prediction from DSSV group who included all the recently published STAR inclusive jet and dijet $A_{LL}$ results at $\sqrt{s} = $ 200 GeV shows $\int_{0.01}^{1}\Delta g(x) = 0.296 \pm 0.108$ at $Q^2 =$ 10 $\textrm{GeV}^2$ \cite{dssv2019}. However large uncertainties of $\Delta g(x)$ still exist at $x < 0.01$. To explore the low $x$ gluon polarization that is not well constrained by current available experimental data, we need to increase $\sqrt{s}$ or extend $\eta$ forward.

\begin{align}
x_1 & = \frac{1}{\sqrt{s}}(p_{T,3}e^{\eta_3} + p_{T,4}e^{\eta_4}) \label{eq:x1}
\\
x_2 & = \frac{1}{\sqrt{s}}(p_{T,3}e^{-\eta_3} + p_{T,4}e^{-\eta_4}) \label{eq:x2}
\\
|cos\theta^{\ast}| &= tanh \frac{|\eta_3 - \eta_4|}{2} \label{eq:theta}
\end{align}

In the year 2012, STAR recorded data from 82 $pb^{-1}$ of longitudinally polarized $pp$ collisions at $\sqrt{s} = 510$ GeV, with average beam polarizations for two beams, 54\% and 55\% respectively \cite{polnote}, and $R$ varing from 0.9 to 1.1. The electro-magnetic calorimeter based jet patch triggers, JP0, JP1 and JP2, are optimized to sample three different ranges over jet $p_T$ with thresholds set at 5.4, 7.3 and 14.4 GeV/c. Jets are reconstructed with charged tracks and electro-magnetic towers using the anti-$k_T$ algorithm with the parameter $R = 0.5$ \cite{antikt}.

An off-axis cone method adapted from the ALICE experiment at the LHC \cite{cones} is applied to correct the jet transverse energy due to underlying event contributions. It collects particles inside two cones centered at $\pm \frac{\pi}{2}$ away from the jet in $\phi$ and at the same jet $\eta$. The correction $dp_T$ is taken as $dp_T = \hat{\rho} \times A$, where $\hat{\rho}$ is the averaged energy density of the two off-axis cones and $A$ is the jet area. This method samples the $\eta$ dependence of the underlying event activities.

\begin{figure}[h]
\begin{center}
\includegraphics[width=0.5\columnwidth]{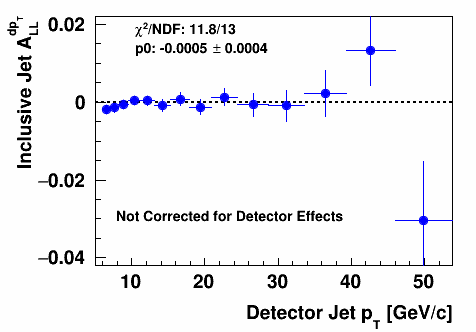}
\end{center}
\caption{\label{fig:aLLdpt}Underlying event correction asymmetry $A_{LL}^{dp_T}$ vs. detector jet $p_T$ measured from data together with a constant fit \cite{run12aLL}.}
\end{figure}

To study its contribution to jet $A_{LL}$, we measure the longitudinal double-spin $dp_T$ asymmetry, $A_{LL}^{dp_T}$ as in Equation \ref{eq:aLLdpt}. A constant fit through $A_{LL}^{dp_T}$ as a function of jet $p_T$ shows that the underlying event correction is consistent with zero, as in Figure \ref{fig:aLLdpt} \cite{run12aLL}. Taking $<dp_T>\times A_{LL}^{dp_T}$ as a shift of the jet $p_T$ due to underlying events, where $<dp_T>$ is the average $dp_T$ regardless of beam helicities, we estimated the potential contribution is at the level of $10^{-4}$, which is assigned as a systematic uncertainty. 

\begin{align}
    A_{LL}^{dp_T} & = \frac{1}{P_1P_2} \frac{<dp_T>^{++} - <dp_T>^{+-}}{<dp_T>^{++} + <dp_T>^{+-}}  \label{eq:aLLdpt}
\end{align}

\begin{figure}[h]
\begin{center}
\includegraphics[width=0.4\columnwidth]{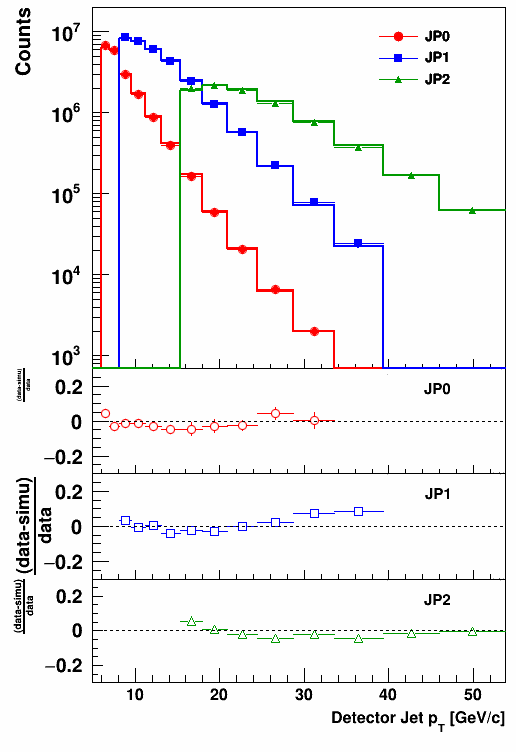}
\end{center}
\caption{\label{fig:jetpt}Jet $p_T$ spectra comparison between data (markers) and embedding (lines) for jets satisfying JP0, JP1 and JP2 triggers separately \cite{run12aLL}.}
\end{figure}

The systematic uncertainties are studied with an embedding sample where simulated hard QCD scattering events are embedded into zero-bias events that are randomly taken during the collisions. The exponent parameter that controls the $\sqrt{s}$ dependent cut-off $p_{T,0}$ in the default Perugia 2012 tune \cite{perugia} was modified to match the simulated $\pi^{\pm}$ spectra with the previously published STAR measurements \cite{pipm2006, pipm2012} from $pp$ collisions at $\sqrt{s} = 200$ GeV \cite{adkins, chang}. Figure \ref{fig:jetpt} shows the excellent agreement between data and simulation for jet $p_T$ spectra for jets satisfying JP0, JP1 and JP2 trigger requirements.

Jets reconstructed from the detector responses in the embedding sample are required to meet the jet patch trigger requirements. Comparing their predicted $A_{LL}$ as a function of jet $p_T$ with the unbiased parton level $A_{LL}$ allows to estimate the trigger bias and reconstruction correction and its uncertainty. The 100 equally probable replicas from NNPDFpol1.1 \cite{nnpdf1.1}, which cover the current uncertainty band of $\Delta g(x)$, results in much more precise estimation of the correction and its uncertainty than the previous measurements at $\sqrt{s} = 200$ GeV.

\begin{figure}[h]
  \begin{minipage}{0.5\linewidth}
    \centerline{\includegraphics[width=0.7\linewidth]{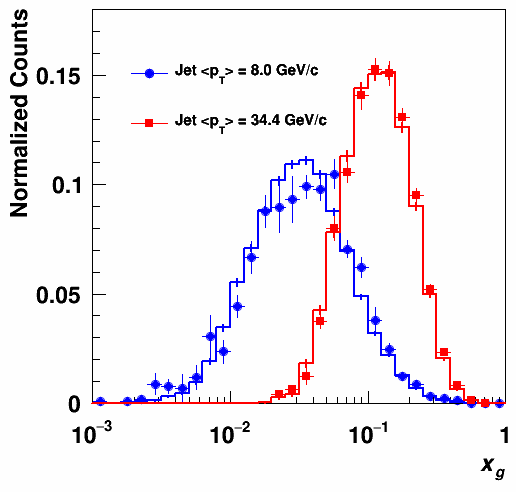}}
  \end{minipage}
  \hfill
  \begin{minipage}{0.5\linewidth}
    \centerline{\includegraphics[width=\linewidth]{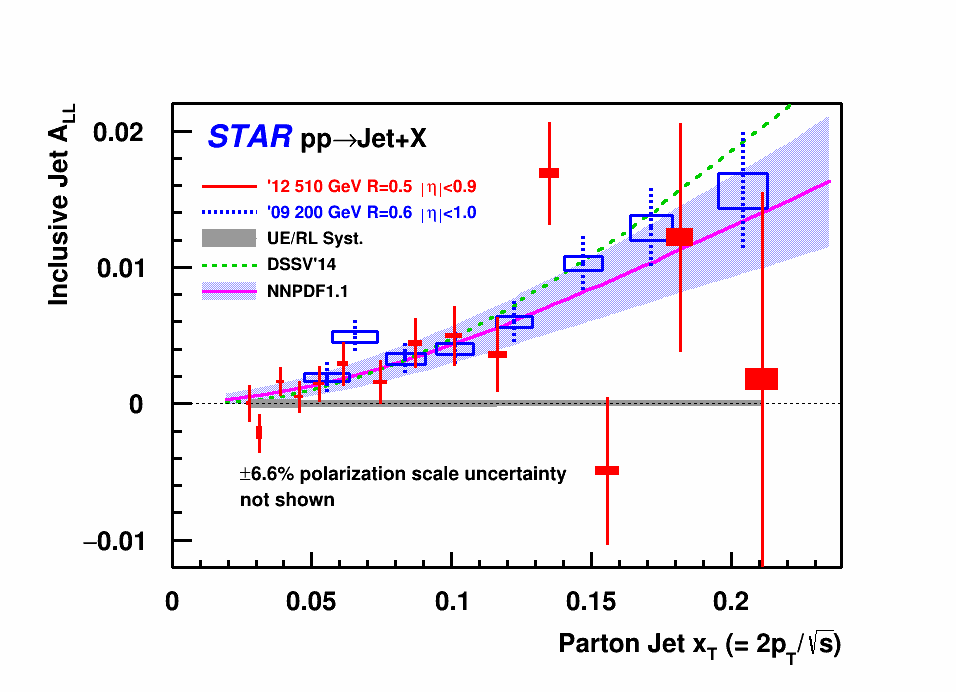}}
  \end{minipage}
  \caption[]{Left panel is the sampled $x_g$ distribution for jets in two $p_T$ bins. Right panel is the inclusive jet $A_{LL}$ vs. parton jet $x_T$ from STAR 2012 510 GeV data (red) and 2009 200 GeV data (blue) together with NLO polarized PDF predictions from DSSV (dashed line) \cite{dssv2014} and NNPDFpol1.1 \cite{nnpdf1.1} models (solid line with shades) \cite{run12aLL}.}
  \label{fig:aLL}
\end{figure}

The STAR 2012 inclusive jet $A_{LL}$ as a function of parton jet $x_T$ at $\sqrt{s} =$ 510 GeV is presented in the right panel of Figure \ref{fig:aLL} \cite{run12aLL}, together with STAR 2009 results at $\sqrt{s} =$ 200 GeV \cite{run9aLL}. Both results agree well in the overlapping $x_T$ region. The new results are also consistent with recent NLO PDF predictions that imply positive gluon polarization \cite{nnpdf1.1,dssv2014}. The 510 GeV results extend measurements to lower $x_T$ which is sensitive to low $x$ polarized gluons. The sensitivities to $\Delta g(x)$ goes to $x$ as low as $\sim$ 0.015, as in the left panel of Figure \ref{fig:aLL} \cite{run12aLL}.

\begin{figure}[h]
  \begin{minipage}{0.5\linewidth}
    \centerline{\includegraphics[width=0.48\linewidth]{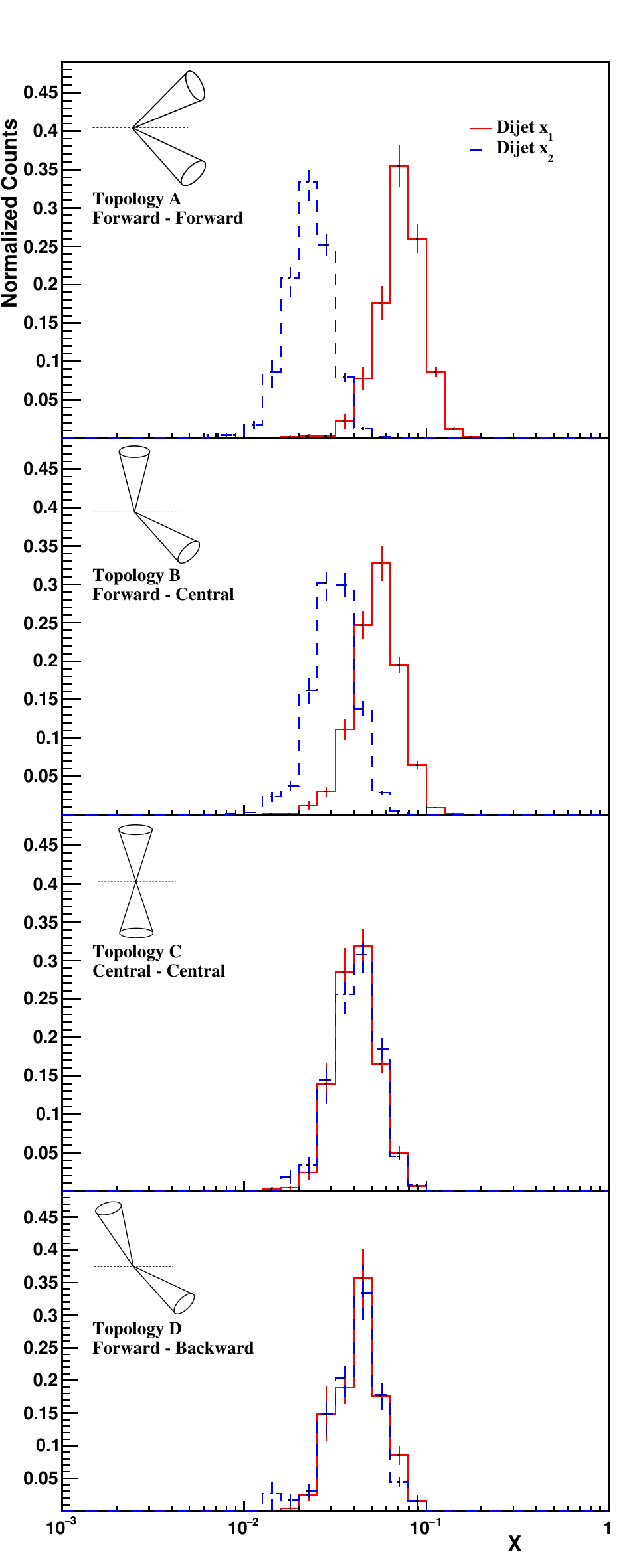}}
  \end{minipage}
  \hfill
  \begin{minipage}{0.5\linewidth}
    \centerline{\includegraphics[width=0.56\linewidth]{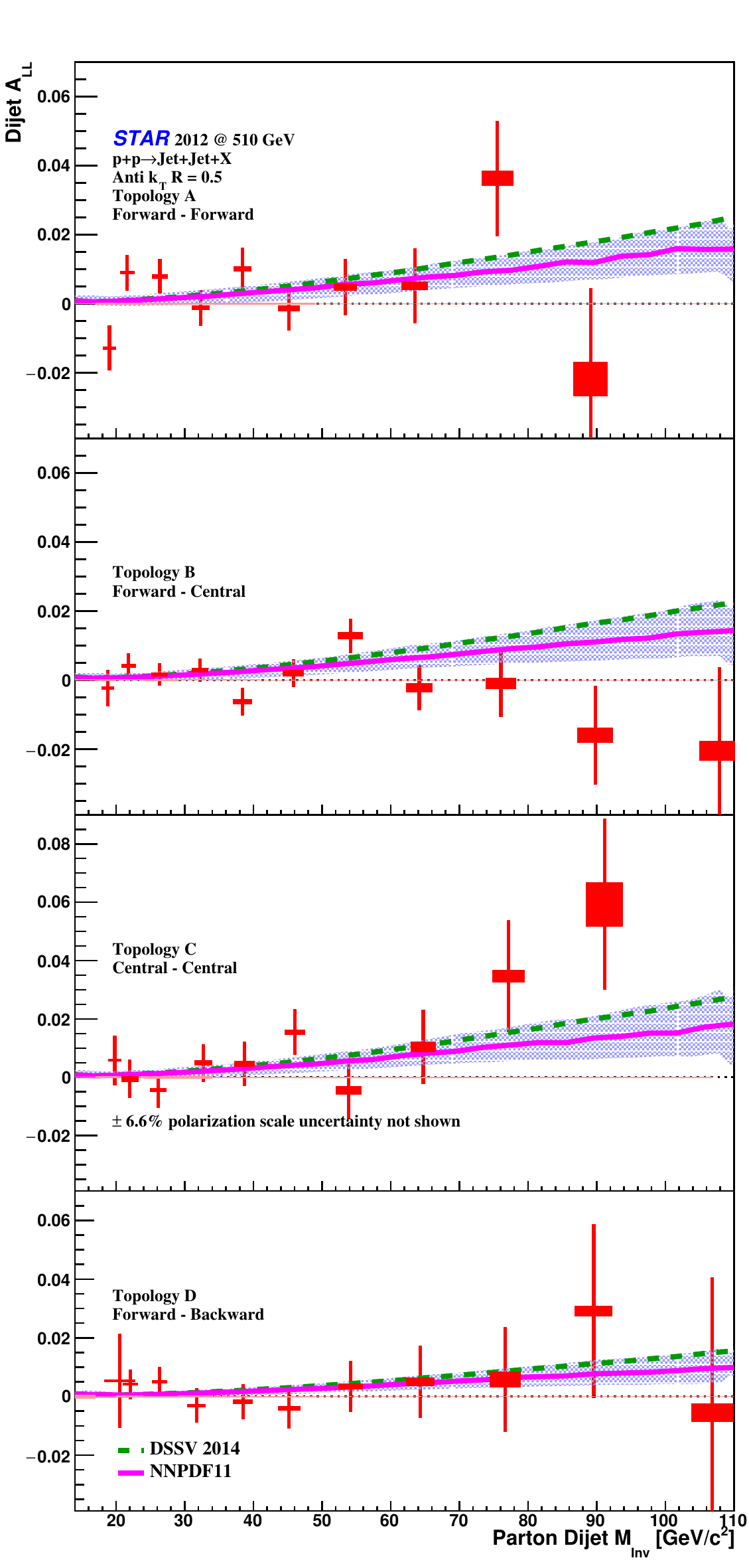}}
  \end{minipage}
  \caption[]{\label{fig:aLLdj} Left panels show the sampled $x_1$  and $x_2$ distributions by dijets for four topologies with invariant mass, $M_{inv} =$ 17 $-$ 20 $\textrm{GeV}/c^2$. Right panels show the dijet $A_{LL}$ vs. parton dijet $M_{inv}$, for four topologies from STAR 2012 data (red) at $\sqrt{s} =$ 510 GeV together with NLO polarized PDF predictions from DSSV (dashed line) \cite{dssv2014} and NNPDFpol1.1 \cite{nnpdf1.1} models (solid line with shades) \cite{run12aLL}.}
\end{figure}

The dijet events require the opening angle $\Delta \phi > \frac{2\pi}{3}$ and the asymmetric $p_T$ cut, $p_{T,3} >$ 6 GeV/$c$ and $p_{T,4} >$ 8 GeV/$c$, for the two jets. The combinations of the unfolded $x_1$ and $x_2$ constrain the shape of $\Delta g(x)$. The partonic $\hat{a}_{LL}$ depends on $cos\theta^{\ast}$.  Therefore we proposed four $\eta$ 
topology binnings as in Table \ref{tbl:top}. As expected, we see the difference in measured dijet $A_{LL}$ for four $\eta$ topologies, as in the right panel of Figure \ref{fig:aLLdj}. The sampled $x_1$ and $x_2$ are much narrower than the sampled $x_g$ by inclusive jets, as in the left panel of Figure \ref{fig:aLLdj} \cite{run12aLL}.

\begin{table}
\caption{\label{tbl:top}Definition of four topologies, forward-forward, forward-central, central-central, and forward-backward for dijets \cite{run12aLL}.}
\begin{center}
\begin{tabular}{llll}
\br
Topology &Description&Regions of $\eta_3$ and $\eta_4$\\
\mr
  A & Forward-Forward& 0.3 $ < |\eta_{3,4}| < $ 0.9, $\eta_3 \cdot \eta_4 >$ 0\\
  B & Forward-Central& $|\eta_{3,4}| < $ 0.3, 0.3 $ < |\eta_{4,3}| < $ 0.9\\
  C & Central-Central& $|\eta_{3,4}| < $ 0.3 \\
  D & Forward-Backward& 0.3 $ < |\eta_{3,4}| < $ 0.9, $\eta_3 \cdot \eta_4 <$ 0\\
\br
\end{tabular}
\end{center}
\end{table}

The preliminary results for the inclusive jet and dijet $A_{LL}$ measurements from STAR 2013 510 GeV $pp$ collisions were released in 2018 \cite{run13aLL}. The integrated luminosity is about four times larger than that of the 2012 data set, however the dijet jet patch triggers were introduced to favorably capture dijet events. The same procedure has been applied in the 2013 inclusive jet $A_{LL}$ measurements. Both results agree with each other. We are finalizing the systematic uncertainties before the future publication.

\section{Other measurements and STAR forward upgrade}
The neutral pions, $\pi^{0}$, are reconstructed from the $\gamma$ decay in the forward meson spectrometer. The measured $\pi^{0}$ $A_{LL}$ results at $\sqrt{s} =$ 510 GeV from STAR 2012 and 2013 data, divided into two $\eta$ ranges, 2.65 $< \eta <$ 3.15 and 3.15 $< \eta <$ 3.90, are very small, less than $5 \times 10^{-3}$. The forward $\eta$ allows to access polarized gluons at $x$ in the order of $10^{-3}$ \cite{fmspi0}. 

The STAR forward upgrade has been fully approved and funded in time for the RHIC 2022 run. It features a forward calorimeter system and a forward tracking system at 2.5 $< \eta <$ 4.0. The calorimeter includes a hadron calorimeter and an electro-magnetic calorimeter. Silicon disks and small thin gap chambers will be installed for the forward tracking system. The dijet $A_{LL}$ will be one of the highlighted physics programs for this upgrade, with one or both jets inside the forward region. With both jets inside the forward region at $\sqrt{s} = 510$ GeV, it allows to sample $\Delta g(x)$ at $x$ as low as $10^{-3}$, where the current model predictions show large uncertainties. The STAR forward upgrade will also lay the ground for the future Electron Ion Collider \cite{fwdupg}.

\section{Conclusion}
In summary, the inclusive jet measurements at STAR probe the magnitude of the $\Delta g(x)$ over a wide range of $x$. The dijet measurements provide additional constraints on the shape of $\Delta g(x)$. The first measurements of inclusive jet and dijet $A_{LL}$ at $\sqrt{s} = 510$ GeV are sensitive to gluons at $x \sim$ 0.015. The results are consistent with current model predictions that imply positive gluon polarizations over $x >$ 0.02. The STAR forward upgrade will play an important role in exploring the gluon polarizations at $x$ near $10^{-3}$, which is loosely constrained by the current world data.

\end{document}